# The Rise and Fall of Anomalies in Tetrahedral Liquids.

Waldemar Hujo,[1] B. Shadrack Jabes,[2] Varun K. Rana,[2] Charusita Chakravarty[2] and Valeria Molinero[1]

[1]*Department of Chemistry, University of Utah, Salt Lake City, UT 84112-0850, USA*
[2]*Department of Chemistry, Indian Institute of Technology-Delhi, New Delhi 110016, India*

Corresponding Authors Information:

V. Molinero, e-mail: Valeria.Molinero@utah.edu; phone: 1-801-585-9618.

C. Chakravarty, e-mail: charus@chemistry.iitd.ernet.in; phone: 91-11-2659-1510.

**Abstract**. The thermodynamic liquid-state anomalies and associated structural changes of the Stillinger-Weber family of liquids are mapped out as a function of the degree of tetrahedrality of the interaction potential, focusing in particular on tetrahedrality values suitable for modeling C, $H_2O$, Si, Ge and Sn. We show that the density anomaly, associated with a rise in molar volume on isobaric cooling, emerges at intermediate tetrahedralities (e.g. Ge, Si and $H_2O$) but is absent in the low (e.g. Sn) and high (e.g. C) tetrahedrality liquids. The rise in entropy on isothermal compression associated with the density anomaly is related to the structural changes in the liquid using the pair correlation entropy. An anomalous increase in the heat capacity on isobaric cooling exists at high tetrahedralities but is absent at low tetrahedralities (e.g. Sn). Structurally, this heat capacity anomaly originates in a sharp rise in the fraction of four-coordinated particles and local tetrahedral order in the liquid as its structure approaches that of the tetrahedral crystal.

# 1 Introduction

Liquids in which the nature of the underlying interactions favours local, short-range tetrahedral order around the particle centers are referred to as tetrahedral liquids.[9] The four-coordinated local order is distinct from the icosahedral order characteristic of simple liquids for which the random, close-packed hard-sphere models are a good zeroth-order approximation.[37, 54] The range of interatomic interactions that can give rise to tetrahedral local order is quite diverse. The best-known tetrahedral liquid is water, where the anisotropic hydrogen-bonding of the molecules gives rise to the liquid-state network.[27] Other examples include ionic melts with suitable radius ratios, such as $SiO_2$, $BeF_2$ and $GeO_2$, and the periodic group IV elements C, Si and Ge and Sn. Despite the diversity of underlying interactions, tetrahedral liquids show a number of common thermodynamic, structural and dynamical behaviours that are anomalous when compared to simple liquids. These can include liquid-state thermodynamic anomalies such as density extrema and a raise in the heat capacity on cooling, polyamorphism, the plausibility of phase transitions between metastable amorphous phases, extensive crystalline polymorphism and a negatively sloped region of the melting curve of the tetrahedrally coordinated crystals.

Interestingly, the full set of anomalies is not present for every tetrahedral liquid. For example, density anomalies have been experimentally observed in water and silica, and in simulation for silicon and beryllium fluoride but not in tin or carbon.[5, 8, 11, 34, 66, 88, 93] Polyamorphism – the existence of distinct low-density and high-density glasses of the same substance – has been reported for germanium, silicon, silica, germania and water.[15, 20, 21, 49, 53, 71, 72, 75] The diffraction patterns of the low-density amorphous phase of water (LDA), silicon (a-Si) and germanium (a-Ge) indicate that they all have the same structure that has been characterized as a random tetrahedral network.[19] Similar structure, although with a non-negligible fraction of three-coordinated sites, was found for amorphous carbon.[29, 33] A high-density amorphous phase of carbon has not yet been found. The main question that arises from these results is whether there is a unifying framework that explains the variability of the anomalies along the series of tetrahedral liquids.

A general approach for understanding the relationship between structure and thermodynamics of liquids, regardless of the nature of underlying interactions, is to define suitable order metrics for the short-range order.[83, 90] In simple liquids and glasses, the local order is characterized using order metrics based on the nature and extent of pair correlations. In the case of tetrahedral liquids, an additional short-range order metric sensitive to local orientational order is required.[30, 88, 89] The anomalous thermodynamic and transport properties of tetrahedral liquids can be understood in terms of a competition between pair order and tetrahedral order. In this work we study the evolution of thermodynamic and structural anomalies as the degree of tetrahedrality is tuned within the Stillinger-Weber (SW) family of liquids. SW is a monatomic potential that consists of a sum of two and three-body interactions between the particles, $E = \phi_2 + \lambda \phi_3$ (for details, see eq. 1 in Methods). The two-body potential $\phi_2$ favours dense packing of particles, while the three-body potential $\phi_3$ adds an energetic penalty to configurations of triplets for which the angle deviates from the tetrahedral angle 109.5°. The three-body penalty encourages low-density tetrahedral configurations and the tetrahedral parameter $\lambda$ controls the strength of tetrahedrality of the potential. Tuning of the relative weights of the two-body and three-body terms in the potential results in a rich phase behavior. The stable crystalline structure at $p = 0$ is the four-coordinated diamond lattice for high $\lambda$ and the eight-coordinated body-centered cubic lattice for low $\lambda$, with a narrow region between them, from $18.2 < \lambda < 18.7$, for which the six-coordinated β-tin is the lowest energy crystal.[59]

The SW potential is particularly appropriate for understanding the effect of decreasing tetrahedrality with atomic number on the phase diagram of the group IV elements. Originally developed for silicon ($\lambda$=21),[84] SW has been parameterized to model germanium ($\lambda$=20)[21] and carbon ($\lambda$=26.2).[16] The narrow region of stability of the β-tin crystal in the phase diagram of this family as a function of $\lambda$, suggests that $\lambda$≈18.5 could model tin.[59] Most recently, the SW potential has been parameterized for water ($\lambda$=23.15).[58] These potentials and their validation are discussed in section 2.1 below. The values of $\lambda$ for these models indicate that the order of tetrahedrality of these substances is C > $H_2O$ > Si > Ge > Sn.

The Stillinger-Weber functional form provides a reasonable representation of a range of experimental systems, though it does not necessarily represent the best or unique potential model for a given substance. The aim of the present work, however, is not to dwell on the quantitative agreement between these models and each of these tetrahedral substances, but to elucidate whether and to what extent the range of anomalous thermodynamic behaviour experimentally observed in tetrahedral liquids can be explained – and predicted when not available – in terms of variations in the degree of tetrahedrality imposed by the underlying interactions. In this effort to define such an underlying common framework for tetrahedral liquids, we investigate the evolution of the density, pair entropy, tetrahedral order and heat capacity within a family of Stillinger-Weber liquids as a function of temperature and tetrahedrality under zero pressure conditions. This work provides a structural basis for understanding why some of the anomalies (such as those in density) are most pronounced at intermediate tetrahedralities while others (e.g. heat capacity) exist in both intermediate and high tetrahedrality regimes.

## 2 Models and methods

### 2.1 Models.

The tetrahedral substances were modeled with the Stillinger-Weber (SW) potential,[84] which consists of a sum of pairwise $\phi_2$ and three-body $\phi_3$ contributions:

$$E = \sum_i \sum_{j>1} \phi_2(r_{ij}) + \sum_i \sum_{j\neq1} \sum_{k>j} \phi_3(r_{ij}, r_{ik}, \theta_{ijk})$$

$$\phi_2(r_{ij}) = A\varepsilon \left[ B\left(\frac{\sigma}{r_{ij}}\right)^4 - 1 \right] \exp\left(\frac{\sigma}{r_{ij} - a\sigma}\right) \quad (1).$$

$$\phi_3(r_{ij}, r_{ik}, \theta_{ijk}) = \lambda\varepsilon (\cos\theta_{ijk} - \cos\theta_0)^2 \exp\left(\frac{\gamma\sigma}{r_{ij} - a\sigma}\right) \exp\left(\frac{\gamma\sigma}{r_{ik} - a\sigma}\right)$$

We used the parameters of the original SW silicon potential, $A = 7.049556277$, $B = 0.6022245584$. $\gamma = 1.2$, $a = 1.8$, $\theta_o = 109.47°$ varying the weight of the three-body term, the tetrahedral parameter $\lambda$, from 18 and 30 to represent a wide range of tetrahedral substances. The simulations were performed using the mass $m = 18$ a.m.u., distance $\sigma = 2.3925$ Å and energy $\varepsilon = 6.189$ kcal mol$^{-1}$ of the monatomic water model mW.[58] The simulation conditions are indicated in mW water units; the results, however, are presented in reduced units (e.g. $T^* = k_B T/\varepsilon$, $\rho^* = \rho\sigma^3$).

The SW potential has been parameterized for silicon, germanium, carbon and water. SW silicon has been extensively validated and shown to reproduce the melting temperatures, slope of the melting curves and polyamorphism of silicon.[17, 26, 48, 49, 59, 77, 91, 92]. SW germanium with $\lambda = 20$ was developed to reproduce the melting temperature, enthalpy of melting and density of the diamond Ge crystal and was shown to reproduce the structure of the liquid.[21] An almost identical SW Ge potential with $\lambda=19.5$ was later shown to produce good agreement with the experimental temperature and enthalpy of "melting" of the amorphous solid a-Ge.[70] The four-coordinated structure of a-Ge, however, cannot be obtained directly by cooling liquid Ge in simulations, as it becomes too viscous –a glass for the time scales accessible to simulations- already before completing the structural transformation to the four-coordinated amorphous phase.[21, 59, 70] We infer a tetrahedral parameter λ between 18.2 and 18.6 for Sn from the phase diagram of the SW family as a function of the tetrahedral parameter λ at $p = 0$, that shows a narrow λ range for which β-tin (the ground state crystal for Sn) is the most stable phase of this family.[59] The SW model for water, the monatomic water model mW, has been shown to reproduce the structures of liquid water, ice, low-density amorphous ice and clathrates and the phase-transformations between them, as well as the thermodynamic and dynamic anomalies of water.[39-43, 58, 60-63] The locus of the thermodynamic anomalies in mW water is displaced by about 30 K down with respect to the values in the experiments.[58, 61, 63] The SW model of carbon of Ref. [16] was developed from ab-initio calculations to reproduce bonds and angles between C atoms and shown to reproduce the elastic constants and bulk modulus of bulk diamond. To the best of our knowledge the SW C potential has not been used to compute the phase diagram of carbon or the properties of its liquid or amorphous phases. We note that in the parameterization of carbon in terms of the SW potential in Ref. [16] the values of some constants in equation 1 are $A = 5.378$, $B = 0.593$, $\gamma = 1.055$ and $a = 1.845$ which are slightly different from the ones we use in this study. In the present work, the SW model potential with the parameter values of Ref. [16] is denoted as carbon or C. The SW potential with the same parameters as Si, but with a tetrahedral parameter λ equal to 26.2 is referred to as carbon* or C*. By comparing the results for the SW liquids of this study and carbon, we find that C would have a tetrahedral parameter λ larger than 30 if the other parameters are those of the SW family studied in this work (see section 3.1), therefore the tetrahedral strength of carbon is significantly higher than for the potential here called C*. Crystallization of the group IV elements has hindered the accurate determination of the loci of the anomalies in these elements, as the anomalies tend to be the most pronounced in the deeply supercooled region. In the case of carbon, its high melting temperature poses an additional experimental challenge for the characterization of the liquid state. Therefore, many of the predictions on the presence or absence of anomalies of these liquids are substantiated by measurements in the available temperature ranges and/or the results of simulations with a variety of classical and quantum mechanical models.

**2.2 Simulation methods.**

Molecular dynamics (MD) simulations were performed using LAMMPS.[67] The simulation cells consisted of either 4,096 or 512 particles in a cubic simulation cell with periodic boundary conditions. The velocity Verlet algorithm was used to integrate the equations of motion with a time step of 5 fs (in mW water units). All simulations were evolved at $p = 0$, in the isothermal isobaric ensemble ($NpT$). Temperature and pressure were controlled using the Nose-Hoover thermostat and barostat with relaxation times of 1.0 and 10 ps, respectively. The set of liquids where cooled at constant rate of 1, 5 and 10 K/ns from which it was determined that the lowest rate that does not result in crystallization of the samples with $\lambda > 21$ was 10 K/ns for the larger system. The formation of crystals was monitored with the CHILL algorithm.[60] Quenching simulations with constant cooling rate 10 K/ns were used for the analysis of the evolution of the properties with temperature for the larger system. Quenching simulations for the smaller system at 1 K/ns and 5 K/ns produced identical temperature-dependent structural properties. The mean square displacements (MSDs) for the atoms were monitored for up to 1 ns. The glass transition temperature ($T_{glass}$) was defined to be the temperature for which no diffusive regime could be detected in the 1 ns period. This is a convenient definition for the cooling rates employed in this study. The $T_{onset}$ temperature was defined as the temperature at which the onset of cooperative or caging effects in the dynamics are signaled by an inflexion point or a plateau region in the log-log plot of the MSD versus time. The 1 ns runs used in this study are adequate for convergence of structural properties at temperatures above $T_{glass}$. We note that care with equilibration is required for temperatures below $T_{onset}$, specially for liquids with tetrahedral parameter in the range 17.5 < λ < 20.5, which are excellent glassformers in simulations. A previous study shows that long runs, for some conditions two orders of magnitude longer than in the present study, may be required in order to obtain reliable estimates of diffusivities for these systems.[59]

**2.3 Property calculation.**

The *melting temperatures* were computed through the direct phase coexistence method.[98] Starting from simulation cells of cubic diamond crystals with approximate dimensions 50 Å × 25 Å × 25 Å, we first equilibrated them for each value of λ and then melted the slab with $x$ < 25 Å, and followed the protocols of Ref. [42] to determine $T_m$ of each substance (defined by its value of λ) with an uncertainty of 1%. The *density* $\rho^* = N\sigma^3/V$ and *enthalpy* $H^* = (E+pV)/\varepsilon$ of the liquids were computed as running averages over the cooling trajectories. The reduced constant pressure heat capacities $C_p^* = \left(dH^*/dT^*\right)_p$ were computed through finite differentiation of $H^*(T^*)$ after the latter were smoothed through running averaging to reduce the noise in the derivative, as in Ref. [58].

The average *tetrahedral ordering* of the 4-closest neighbors around each particle was measured through the order parameter $q_{tet}$,[31] defined for each particle as

$$q_{tet} = 1 - \frac{3}{8}\sum_{i=1}^{3}\sum_{j=i+1}^{4}\left(\cos\theta_{ikj} + \frac{1}{3}\right)^2 \qquad (2)$$

where $q_{ikj}$ is the angle subtended between the central particle $k$ and two of its 4-closest neighbors. The average value of the tetrahedral order parameter, $\langle q_{tet}\rangle$, corresponds to the average of $q_{tet}$ over the N particles. The value of $\langle q_{tet}\rangle$ is one for a perfect tetrahedral crystal where all $q_{ikj}$ = 109.47° and zero for a system in which the distribution of these angles is random. The first coordination shell of the particles in the liquid was defined from the first minimum of the radial pair distribution function of the liquids. The *fraction of four-coordinated particles $f_4$* corresponds to the fraction of particles with exactly four neighbors within the first coordination shell.

The total entropy (*S*) of a fluid can be written as the sum of an ideal contribution, $S_{id}$, and contributions $S_n$ from n-body correlations i.e. $S = S_{id} + S_2 + S_3 + ... = S_{id} + S_e$ where $S_e$ is the excess entropy of the fluid relative to the ideal gas at the same temperature and density. Note that this definition of the excess entropy of the liquid is different from that frequently used in the literature on glasses where the excess entropy of the liquid is measured relative to the crystal. The *pair entropy* (*$S_2$*) is the contribution to the total entropy of a fluid due to pair correlations and is defined using the radial distribution function, *g(r)*, as [14, 36, 65]

$$\frac{S_2}{Nk_B} = -2\Pi\rho\int_0^\infty \{g(r)\ln g(r) - g(r) + 1\}r^2 dr \quad (3)$$

When evaluating $S_2$ from simulations, the integration was performed until a cut-off distance corresponding to half the simulation box length. In the case of simple liquids, the pair correlation contribution accounts for 85-90% of the excess entropy. In the case of tetrahedral liquids, the three-body contributions are expected to be more significan. Nevertheless, previous studies on water, $SiO_2$ and $BeF_2$ suggest that, even in these tetrahedral systems, the pair entropy remains the dominant contribution.[3, 80] The pair entropy therefore provides a reasonable structural estimator for the excess entropy (*$S_e$*) of a fluid.

## 3 Results.

### 3.1 Density

We first consider the behaviour of the density as a function of temperature $\rho(T)$ along the $p = 0$ isobar for the systems with different tetrahedral parameter $\lambda$ in order to identify thermodynamic anomalies associated with tetrahedral liquids, particularly the existence of density maxima and minima as well as possible signatures of a sharp structural transformation in the liquid. These characteristic thermodynamic features are then mapped out in the temperature-tetrahedrality plane and their sensitivity to cooling/heating protocols and equilibration issues is discussed.

The decrease of the melting temperature with pressure is an unusual characteristic that ice shares with the diamond structure crystals of silicon, germanium and tin. The experimental slope of melting of diamond carbon, on the other hand, is positive. This reflects a change in sign of the volume of melting $\Delta V_m$, as $(dp/dT)_{coex} = \Delta V_m/\Delta S_m$. Consistent with the experimental results, the volume of melting of the tetrahedral crystals increases with tetrahedrality $\lambda$ of the substance, passing through zero at $\lambda = 24.4$, between water and carbon, for the SW family of this study.[58]

The three-body contribution to the energy is zero in a perfectly tetrahedral crystal, thus the energy and volume of the crystal are relatively insensitive to variations in $\lambda$. This implies that the reversal to the "normal" situation of a crystal denser than the liquid at high tetrahedrality is entirely due to the increase of the volume of the liquids with $\lambda$. This is evident in Figure 1, which presents the density of the family of tetrahedral liquids and the diamond cubic crystal as a function of temperature. Carbon with the parameters of Ref. [16] is more tetrahedral than C*, the potential with $\lambda = 26.2$ of the family of this study, and produces a very low-density liquid (about half of the density of the crystal) in qualitative agreement with the experiments that indicate that the density of liquid carbon is even lower, about one third of the density of diamond.[79] Most noteworthy in Figure 1 is the lack of density anomaly for both low and high values of tetrahedral parameter $\lambda$. The density anomaly, a density maximum followed by a density minimum at lower temperatures, is only observed for the liquids with tetrahedral parameter in the range $20 < \lambda < 26.4$. The SW models predict that the range of tetrahedrality that produces anomalous density at $p = 0$ encompasses germanium, silicon, water and silica, but excludes tin and carbon. For these liquids which show a density anomaly, a liquid transformation temperature ($T_L$) can be identified as the locus of maximum thermal expansivity, between the temperature of maximum density (TMD) and the temperature of minimum density (TmD). It is interesting to note that liquids with $\lambda > 26.4$, beyond the region of density extrema, still display an inflection point of the density on cooling. In section 3.3 we show that the inflection in $\rho(T)$ of the highly tetrahedral liquids ($\lambda > 26.4$) is associated with a sharp increase in tetrahedral order and fraction of four-coordinated particles in the structure of the liquid and results in an anomalous increase in the heat capacity of the highly tetrahedral liquids, even if these do not present a density anomaly.

Figure 2 presents the $T$-$\lambda$ phase diagram of the family of liquids at $p = 0$. The tetrahedrally coordinated crystal is the stable low temperature phase for $\lambda > 18.7$.[59] The stable crystals for liquids with with $\lambda < 18.7$ ($\beta$-tin, body centered cubic (bcc) and face centered cubic (fcc), in order of decreasing $\lambda$[59]), have higher density than the liquids from which they form: notice the sudden increase in density of $\lambda = 16$ in Figure 1 as the liquid crystallizes to bcc. In general, liquids approach the structure of the stable crystal on cooling, therefore liquids with $\lambda < 18.7$ display a monotonous increase in density on cooling. Liquids with high values of $\lambda$ are, similarly, less dense than the corresponding tetrahedral crystals and their density increases monotonically on cooling. The loop defined by the *TMD* and *TmD* lines encloses the region of density anomaly at $p = 0$ for this family of liquids. The density anomaly in the SW family is correlated with the existence of a stable crystal phase of density lower than the liquid. The locus of states corresponding to a liquid transformation temperature, that in section 3.4 we show it also corresponds to a maximum in the heat capacity, runs between the TMD and TmD loci and extends to high tetrahedralities for which the density anomaly disappears.

We briefly discuss equilibration issues and the sensitivity of the $T$-$\lambda$ phase diagram to cooling/heating protocols. The densities presented in Figure 1 were computed over cooling simulations. SW systems with $17.5 < \lambda < 20.25$ are good glass formers in the time scales accessible to molecular dynamics simulations and therefore equilibration cannot be achieved at low temperatures.[59, 62, 77] For example, the root mean square displacement over 1 ns run at the temperature of maximum change in the density (our definition of $T_L$) ranged from about $1\sigma$ for $\lambda = 20$ (Ge) to approximately $6\sigma$ for $\lambda = 26.2$ (C*). The liquid with $\lambda = 20$ could not be properly equilibrated at $T <$ TMD at the cooling rates of this study and the regions of its density anomaly is subject to some error. Another consequence of the high viscosity of the liquids around $\lambda \approx 20$ is that the glass formed by the hyperquenching simulations does not reach the four-coordinated structure characteristic of a-Ge in experiments,[21, 59, 70] resulting in a higher density glass than the glasses for silicon and water, while the experimental diffraction patterns of the low-density amorphous water, silicon and germanium indicate that all of them are tetrahedrally coordinated.[19] The liquids with $\lambda > 21$, on the other hand, reach $T_L$ with quite low viscosity and can be properly equilibrated down to about that temperature at which their viscosity increases as they become mostly four-coordinated (see section 3.3 below). The use of slow cooling rates or long isothermal runs to equilibrate these liquids with a tetrahedral crystal ground state at $T < T_L$ results in spontaneous crystallization. Crystallization is also observed for SW for $\lambda = 16$ that forms a bcc crystal and does not present a density anomaly. In this case, the liquid can be equilibrated down to the very low temperature at which it crystallizes to bcc. In summary, the density anomaly does not exist in the SW family of this study for $\lambda > 26.2$ on the high end and conclusively there is no anomaly on the region where the liquids crystallize to crystal with coordination higher than four.

**3.2 Entropy.**

The density anomaly corresponds to the condition that the thermal expansion coefficient, $\alpha$, is greater than zero. Using Maxwell's relations, this can be shown to be equivalent to the condition that the entropy $S$ increases under isothermal compression,

$$\left(\frac{\partial S}{\partial \rho}\right)_T = V^2 \left(\frac{\alpha}{\kappa_T}\right) . \tag{4}$$

Since $S = S^{id} + S^e$, and $S^{id}$ is a monotonically decreasing function of density, we expect that the density anomaly should be associated with a rise in excess entropy on isothermal compression. It has been shown that the excess entropy anomaly is a necessary condition for observing water-like thermodynamic and mobility anomalies.[2-7, 32, 38, 80, 81] Approximating the excess entropy by the pair entropy should lead to the density anomaly being accompanied by a pair entropy anomaly, defined as

$$\left(\frac{\partial S_2}{\partial \rho}\right)_T > 0 . \tag{5}$$

The presence of a pair entropy anomaly corresponding to a set of state points with a positive correlation between $S_2$ and density therefore provides a structural indicator of anomalous behaviour, as shown previously for liquids with water-like anomalies, such as water,[4, 25, 44] silica,[81] beryllium fluoride,[1, 2, 6, 7] germania[38] and two-scale ramp fluids.[22, 23, 32, 56, 57, 94, 95, 97] Figure 3 shows that the liquids with $\lambda=20$ (germanium), 21 (silicon) and 23.15 (water) present such a regime of anomalous pair entropy behaviour. The pair entropy anomaly for $\lambda=20$ is weak, in accordance with the weak density anomaly discussed in section 3.1, and is subject to the same qualifiers with regard to equilibration of highly viscous, intermediate tetrahedrality systems. In the low tetrahedrality limit, the liquid with $\lambda=18.2$ (that we expect to be a reasonable representation for Sn) does not present density or pair entropy anomalies and $S_2$ decreases monotonically with density. While the pair entropy anomalies also disappear in the limit of very high tetrahedrality, the behaviour of the $S_2(\rho)$ curves is qualitatively different from the low tetrahedrality limit. Instead of the non-monotonic behaviour seen at intermediate tetrahedralities, as temperature is lowered, $S_2$ initially decreases smoothly until a sufficiently low temperature is reached when it drops abruptly without any significant change in density. In section 3.3 we show that this decrease in entropy is associated to the increase in tetrahedral order in the liquids. The systems with intermediate $\lambda$ show a similar drop in $S_2$ after going through a regime of anomalous density behaviour. The maxima in thermal expansivity discussed in section 3.1 occur in this region of steep decrease in the $S_2(\rho)$ curves and coincide with the heat capacity maxima discussed in section 3.4. Thus, the pair entropy provides a very convenient and experimentally accessible link between structural correlations and thermodynamic properties of fluids.

While strong pair correlations dominate the structure of liquids, and therefore the pair entropy is typically the dominant contribution to the excess entropy, it is reasonable to expect that the contribution of the three-body correlations will be important for SW liquids with high tetrahedral parameters $\lambda$. It is therefore important to demonstrate that the qualitative behaviour of the thermodynamic excess entropy ($S_e$) and the pair entropy ($S_2$) is similar for these systems. The above discussion shows the parallel behaviour of the density and pair entropy anomalies. The other important thermodynamic anomaly is the heat capacity peak associated with the liquid transformation temperature. Since $C_p = T(\partial S/\partial T)_P$, then a maximum in the heat capacity suggests a steep change in total entropy $S$ that must be reflected in a steep change in the excess entropy $S_e$. We see a steep drop in the pair entropy in the region of the heat capacity anomaly (presented in 3.4 below) suggesting that the change in pair entropy contributes to in the heat capacity anomaly, even though the pair entropy approximation by itself may not be accurate enough to quantitatively determine the behaviour of the total heat capacity of the liquids. We plan to map out the excess and three-body entropies for this family of liquids in a future study.

### 3.3 Tetrahedral order.

We now consider the evolution of the structure of the SW liquids as a function of temperature. We use two metrics of structure: the fraction $f_4$ of atoms with four neighbors in their first coordination shell and the average value and distribution of the tetrahedral order parameter $q_{tet}$. Figure 4 shows the behaviour of $q_{tet}$, defined by eq. 2., as a function of temperature along the $p = 0$ isobar for this family of SW liquids. The value of $q_{tet}$ for $\lambda = 18.2$ is low and essentially temperature-independent. This result, along with the lack of density anomaly for this potential, points to a lack of thermodynamic anomalies in tin, consistent with the available experimental reports.[66] The small increase in tetrahedrality in this model of liquid tin at low temperatures seems to be a vestige of the structural transformation observed in more tetrahedral fluids, and indicates that, different from simple fluids and in agreement with x-ray analysis of liquid tin in experiments,[28, 64] it has a small degree of anisotropy in the interatomic interactions.

In the liquids that show a density anomaly, decrease in temperature initially results in a relatively slow rise in $q_{tet}$, followed by an intermediate temperature regime where there is a sharp rise in $q_{tet}$ before a plateau value is reached at low $T$. Increasing tetrahedrality results in a slower rise in $q_{tet}$ with temperature, possibly because of the already high degree of tetrahedral order associated with the high-temperature liquid.

To understand the dynamics associated with the temperature dependence of $q_{tet}$, we plot in Figure 4 the temperature for onset of cooperative behaviour ($T_{onset}$) that signals the appearance of a plateau region of the MSD versus time plots of the liquids. The $T_{onset}$ approximately coincides with the temperature at which $q_{tet}$ shows a rapid rise for the intermediate $\lambda$ values. Figure 4 also shows the temperatures $T_{glass}$ for which the MSD show no diffusive behaviour within 1 ns simulation runs. Clearly, the local caging effect is associated with the onset of greater local tetrahedral order. The low-temperature plateau in $q_{tet}$ is reached at low temperatures when diffusive dynamics does not occur on the time scale of the simulations.

To obtain some further structural insight into the disappearance of the density anomaly at high $\lambda$, we plot in Figure 5 the distributions of tetrahedral order $P(q_{tet})$ at equally spaced temperatures for $\lambda=23.15$ (water) and $\lambda=26.2$ (C*, less tetrahedral than carbon). The differences between water, which has a well-defined density anomaly, and C*, which is close to the locus where the anomaly disappears, are immediately apparent. In the case of water, $\lambda=23.15$, the distributions $P(q_{tet})$ have a peak at a high $q_{tet}$ value of about 0.8, together with a strong shoulder at $q_{tet}$ of approximately 0.5. The presence of a strong shoulder in addition to a tetrahedral peak is characteristic of the anomalous regime[30] and suggests facile transitions between local tetrahedral and non-tetrahedral environments. As temperature decreases, the non-tetrahedral shoulder becomes less prominent and eventually vanishes, giving rise to a distribution with a sharp peak at high $q_{tet}$. The transition from a strongly shouldered distribution to a distribution dominated by tetrahedral order is spread over a wide temperature range. Thus the competition between pair and tetrahedral order responsible for the rise in excess entropy with density is apparent. In contrast, for $\lambda=26.2$ the highest temperature $P(q_{tet})$ distribution shows some evidence of a shoulder but this signature disappears over a very small temperature range. Further increase in temperature sharpens the tetrahedral peak of the distribution. Thus, systems of moderate tetrahedrality seem to accommodate a much higher fraction of non-tetrahedral defects at a given temperature and pressure. The fact that the pair entropy and density anomalies seem to be the most pronounced for this regime suggests that these defects correspond to relatively high local coordination numbers. Strongly tetrahedral systems will support a much smaller fraction of such defects, mostly corresponding to coordination numbers lower than four.

As an additional structural metric that can be correlated with the thermodynamic behaviour, we computed the fraction $f_4$ of four-coordinated atoms in the liquids along the quenching simulations. As previously shown for mW water,[61] the behaviour of $f_4(T)$ parallels the behaviour $q_{tet}(T)$: the temperatures where $(dq_{tet}/dT)_{p,\lambda}$ and $(df_4/dT)_{p,\lambda}$ reach a maximum are essentially identical, and also coincide with the $T_L$ determined from the maximum in $(d\rho/dT)_{p,\lambda}$. The line of maximum change in fraction of four-coordinated particles at $p = 0$ is presented in the $T$-$\lambda$ phase diagram of Figure 2. Different from the density anomaly, the existence of a pronounced increase in $f_4$ on cooling is observed up to the highest values of $\lambda$ of this study, as the supercooled liquid approaches the tetrahedral structure of the stable crystal. The liquids that stabilize the bcc crystal, on the other hand, do not display an increase in the fraction of four-coordinated particles on cooling.

### 3.4 Heat capacity.

The constant pressure heat capacity $C_p$ of normal liquids decreases with temperature. Experiments indicate that water and beryllium fluoride present an anomalous increase in the heat capacity in the supercoooled region.[13, 87] An anomalous increase in the heat capacity was also observed in simulations of silica and silicon.[9, 59] We are not aware of experimental determinations of the heat capacity of supercooled liquid carbon, tin or germanium. Figure 6 presents the $C_p$ of the SW family as a function of temperature. Liquids within the glass forming region of the SW family ($17.5 < \lambda < 20.25$[59]) present a weak maximum in $C_p$, that we interpret mostly as arising from the loss of ergodicity during vitrification, although the existence of a density anomaly for $\lambda > 19$ suggests that for $19 < \lambda < 20.25$ there is also a contribution from the structural transformation of the liquid. For liquids with tetrahedral parameter 21 and larger, the heat capacity displays an anomalous increase that results in a sharp maximum in $C_p$ in the supercooled region. This heat capacity anomaly seems to start at the same lower tetrahedrality of the density anomaly but extends well beyond where the TMD and TmD lines merge. The liquids that stabilize a tetrahedral crystal display a marked increase in heat capacity in the supercooled region that is directly associated to the structural transformation monitored by the steep increase in the fraction of four-coordinated molecules and tetrahedral order (Figure 4). The locus of $C_p$ maximum coincides with the locus of maximum change in the fraction of four-coordinated particles $(df_4/dT)_{p,\lambda}$, as shown in Figure 2. The strong coupling of the thermodynamic anomaly and the change in structure of the liquid should not be surprising: the enthalpy of the SW liquids depends on the fraction of tetrahedrally coordinated particles, and thus the increase in heat capacity is associated to the sharp increase in tetrahedral order (or the fraction of four-coordinated particles) with temperature. We note, however, that the fact that the enthalpy of the liquids that crystallize to a tetrahedrally coordinated crystal decreases as the fraction of tetrahedrally coordinated molecules increases does not imply that the heat capacity will increase on cooling. The anomaly observed in the heat capacity points to a *cooperative* increase in the fraction of four-coordinated particles in the supercooled liquids with tetrahedral ground state crystal. Already three decades ago Stanley and Teixeira used a bond percolation model to explain how this cooperativity may arise in water.[82] There is no equivalent rush towards the formation of a mostly eight-coordinated liquid for the SW systems with bcc ground state.

The structural transformation of the supercooled tetrahedral liquids that produces the anomalous increase in the heat capacity is also responsible for the fast nucleation of the tetrahedral crystal. A detailed analysis of the kinetics of crystallization of mW water ($\lambda$ = 23.15) as a function of temperature, demonstrates that at temperatures below the locus of $C_p$ maximum, the crystallization rate of the supercooled liquid is comparable to the time scale required for its relaxation, and the liquid cannot be equilibrated.[62, 63] The fast crystallization observed in the tetrahedral liquids of this study after they cross the peak in the heat capacity and transform into mostly tetrahedrally coordinated amorphous structures, suggests that the limit of metastability reported for water in Refs. [62, 63] may be general to the full class of tetrahedral liquids.

**4 Summary and discussion.**

The similarities in the phase diagrams and liquid-state behaviour of water, many of the Group IV elements and certain $AB_2$ halides and oxides suggests that the tetrahedral topology dictated by the interactions is more important in defining the thermodynamic and structural behaviour of these substances than the specific nature (metallic, covalent or hydrogen-bonded) of their interactions. The Stillinger-Weber systems are ideal for understanding the structural and thermodynamic implications of variable tetrahedrality, specially since a wide range substances, encompassing covalent, semimetals and metals of group IV (C, Si, Ge, Sn), and a hydrogen bonded molecular liquid (water) can be mapped onto this common parametric potential form. In this study, we compute the melting temperatures and the existence and locus in the temperature tetrahedrality ($T$–$\lambda$) plane of the density, heat capacity and structural anomalies of these SW liquids at zero pressure.

The stable, low-temperature crystalline phase for the Stillinger-Weber potential adopts distinct structures on increasing the tetrahedral strength $\lambda$, changing from fcc to bcc, then to β-tin, and finally to tetrahedral. Hyperquenching of the highly tetrahedral SW liquids results in the formation of tetrahedrally coordinated amorphous solids, of which low-density amorphous ice, amorphous silicon, amorphous carbon and amorphous germanium are experimental realizations, although the latter cannot be obtained by cooling of the SW Ge liquid in time scales available to simulations. The tetrahedrally coordinated liquids that vitrify onto these amorphous solids cannot be studied in equilibrium, as they easily crystallize into the tetrahedral crystals. The tetrahedral diamond crystal, stable above $\lambda > 18.7$, shows little change in density with $\lambda$, while the liquid density is very sensitive to the strength of the tetrahedral repulsions, controlled by $\lambda$. The range of tetrahedralities for which the liquid density anomaly exists is similar though not identical to those for which the tetrahedral crystalline state is stable and shows a negative molar volume change on melting.

This work demonstrates that the density anomaly in SW liquids is only observed for tetrahedral parameters λ for which the low temperature tetrahedral amorphous phase has lower density than the high temperature (less tetrahedral) liquid. As a function of increasing tetrahedrality, the density anomaly in the simulations emerges for λ around 19.5 and disappears again at high tetrahedrality, λ=26.4 for the family of this study. Carbon is an example of highly tetrahedral liquids for which we predict that there is no density anomaly. This association of density anomalies with intermediate degrees of tetrahedrality in the $T$-λ plane is analogous to the behaviour reported for water in the $T$-$p$ plane. High pressure, like low tetrahedrality, promotes pair ordering and simple liquid behaviour, attenuating the density anomaly.[9, 11] Negative pressures –that promote tetrahedrality in the structure of liquid water– also result in the loss of the density anomaly,[68] analogous to the behaviour observed for the full range of SW liquids as a function of tetrahedral parameter λ at $p = 0$. This suggests that increasing pressure and decreasing tetrahedrality of the interactions have equivalent effect on the thermodynamics of tetrahedral liquids. The extent to which these two means of decreasing the tetrahedral order in the liquid are quantitatively the same, e.g. whether germanium behaves like compressed silicon, is an interesting question that is left for future research.

An experimentally testable link between structure and thermodynamics is provided by the pair entropy anomaly, which closely tracks the density anomaly and corresponds to a rise in the pair correlation contribution to the liquid entropy on isothermal compression. Since the Stillinger-Weber liquid can be mapped onto several real liquids, construction of pair entropies from experimental radial distribution functions would be of interest. In systems that show a density anomaly, isothermal compression results in a loss of ordering due to pair correlations, in contrast to the behaviour seen in simple liquids. To characterize the accompanying changes in orientational order of the SW liquids, we use local tetrahedral order ($q_{tet}$) and the fraction of four-coordinated molecules ($f_4$). The local order in the liquid structure is dominated by similar coordination numbers as the corresponding crystalline states, but the random network structure accommodates higher defect densities and there are no distinct phase changes as a function of tetrahedrality. At the intermediate tetrahedralities where the density anomaly is observable, the tetrahedral order distributions suggest a significant population of both locally tetrahedral and non-tetrahedral atomic environments over a wide range of temperatures. In conjunction with the negative volume change on melting, this indicates that the non-tetrahedral local environments or defects in these liquids result in a mean coordination number greater than four. As tetrahedrality increases, such high coordination defects become less probable, and the density of the liquid decreases and eventually falls below that of the crystal resulting in a return to expansion of volume on fusion. The relationship between the pair entropy anomaly and the density anomaly may be seen as a consequence of the ability of liquids such as water and silicon to adopt high coordination numbers in response to compression; this ability is lost for strongly tetrahedral liquids such as carbon.

All the SW systems with a tetrahedral ground state crystal show a continuous structural transformation when the liquid undergoes a cooperative increase in four-fold coordination or tetrahedral order on cooling. This structural transformation is reflected in an anomalous increase in the heat capacity on isobaric cooling. The heat capacity anomaly is observed for a wider range of tetrahedralities than the density anomaly. On the low tetrahedrality side, the heat capacity peak becomes smaller and seems to disappear simultaneously with the density maximum. This is accompanied by vitrification of the liquid in the high-temperature structure, as previously reported in Ref. [59]. The vitrification is the result of the limited time scales accessible to simulations; the group IV elements cannot be vitrified by cooling their liquid in experiments around $p = 0$. Different from the density anomaly, the heat capacity anomaly persists up to the highest values of tetrahedrality of this study. We predict that all the tetrahedral liquids with a tetrahedral crystal ground state will present a heat capacity anomaly deeply in the supercooled region. The heat capacity increase is due to a cooperative change in the structure of the liquid, toward a more tetrahedral ordering. This structural transformation closes the gap between liquid and crystal and prevents the equilibration of the supercooled tetrahedral liquids below the locus of the $C_p$ maxima. As argued elsewhere for water,[62, 63] the locus of $C_p$ maximum determines the limit of metastability of the liquid state: at lower temperatures, fast crystallization prevents a proper equilibration of the supercooled liquid phase. Verification of the existence of a limit of metastability for the full class of tetrahedral liquids requires the determination of the characteristic crystallization and relaxation times of the liquid, endeavor that is beyond the scope of this work.

The existence of a first order liquid-liquid transition, (LLT) and a liquid-liquid critical point (LLCP) in water and in silicon has been widely discussed in the literature.[50, 51, 53, 54, 69] A first order liquid-liquid transition that ends in a liquid-liquid critical point located at positive pressure has been proposed for water and, a LLCP at negative pressures has been proposed from simulations of SW silicon.[18, 52, 51, 53, 55, 69, 73, 77, 78, 91, 96] We do not observe direct evidence of a LLT or a LLCP in the simulations of this study, and the unavoidable crystallization of the liquids at temperatures below the $C_p$ maximum suggests that, as discussed above, it is not possible to equilibrate a more tetrahedral low-temperature liquid phase, at least when the size of the simulation cell is much larger than the critical size of the crystal nuclei around the liquid transformation temperature $T_L$. More work is needed to establish whether the low-temperature tetrahedral amorphous phases of any tetrahedral liquid can be properly equilibrated before it nucleates the tetrahedrally coordinated crystal phase.

The present work shows how the thermodynamic anomalies rise and fall as a function of the strength of the tetrahedral repulsion λ for a model potential (SW), and it demonstrates a nesting of the anomalous behavior as a function of tetrahedrality that is reminiscent of the ordering of the anomalies with pressure. The rise and fall of anomalies spanned by the SW family mirrors the one observed for the tetrahedral substances of the group IV and water, therefore we conclude that the variation in anomalies along this real series can be explained in terms of the different strength of tetrahedrality of their intermolecular interactions. The study of the Stillinger-Weber family of liquids suggests that a considerable part of the complex phenomenology of real tetrahedral liquids, with variable degrees and combinations of various anomalies, can be understood as a consequence of differing degrees of tetrahedrality imposed by the underlying interactions. Our results indicate that the commonality in this series of liquids arises from the common topology of the intermolecular interactions, and not from the detailed nature of the intermolecular potential, that in the actual substances modeled by the SW potential span from covalent to hydrogen-bond. A first natural extension is to ask whether it is possible to map all tetrahedral liquids onto a common tetrahedrality scale. Network-forming ionic melts, such as $SiO_2$, $GeO_2$ and $BeF_2$, are obvious systems to consider in this regard. The molar volume change on melting of the cristobalite phase of silica is almost zero, suggesting that the tetrahedrality of silica is intermediate between those of water and carbon and probably just above $\lambda = 24.4$ if mapped onto the SW family. The relationship between tetrahedral and pair correlation ordering in these systems appears to be somewhat different from that seen for Stillinger-Weber liquids,[38] suggesting that an additional variable may be needed to describe the phase behaviour and anomalies of binary network tetrahedral liquids. A second interesting possibility suggested by this study is the design of liquids of different tetrahedrality to control the strength of their anomalies. One route is to consider binary mixtures, such as $GeO_2$-$SiO_2$ and $BeF_2$-$LiF$, where there is compositional control of tetrahedrality.[10, 47, 76] The investigation of mixtures would allow to elucidate whether the composition plays the same role as the pressure and tetrahedral parameter $\lambda$ in tuning the anomalies of the liquid or, as has been observed upon addition to water of solutes such as salts that favour high coordination states, the result is phase separation of liquids with different degree of tetrahedral ordering.[12, 24, 45, 46, 50, 85, 86] Another alternative may be provided by the design of mesoscopic analogues of tetrahedral liquids, such as patchy colloids.[35, 74, 99]

**Acknowledgements.**

We gratefully acknowledge the support by the Arnold and Mabel Beckman Foundation through its Young Investigator Program (V.M.) and the Department of Science and Technology of New Delhi (C.C.). W.H. thanks the German Academic Exchange Service (DAAD) and the exchange program at TU-Braunschweig for supporting his stay at the University of Utah through a student fellowship. B.S.J. thanks the Indian Institute of Technology, New Delhi, for the award of a Junior Research Fellowship. We thank the Center of High Performance Computing at the University of Utah for allocation of computing time.

**Figure 1.** Evolution of the reduced density of the SW liquids and crystals as a function of reduced temperature, computed over quenching simulations with 4,096 particles. The red and black solid lines indicate the density of the liquids of the SW family of this study with tetrahedral order parameters $\lambda$ = 16, 18.5, 19, 19.5, 20, 20.5, 21, 21.5, 22, 22.5, 23, 23.5, 24, 25, 26, 27, 28, 29, 30 and 35. Carbon modeled with the parameters of ref. [16] (dashed line) does not present anomaly and it has a considerably higher tetrahedral strength than C*, the SW liquid of the silicon SW family with $\lambda$ = 26.2. The region of the density anomaly (red curves) encompasses germanium, silicon and water, but excludes tin and carbon. The dashed blue line indicates the density of the diamond cubic (DC) tetrahedral crystal. The sudden increase in density for $\lambda$ =16 is due to the crystallization of the BCC crystal, and the change in the slope of the density of 18.5 and is due to vitrification of the liquid. The locus of maxima in the $C_p(T)$ curves is shown in green; equilibration at temperatures below this line is difficult either due to high viscosity or due to crystallization.

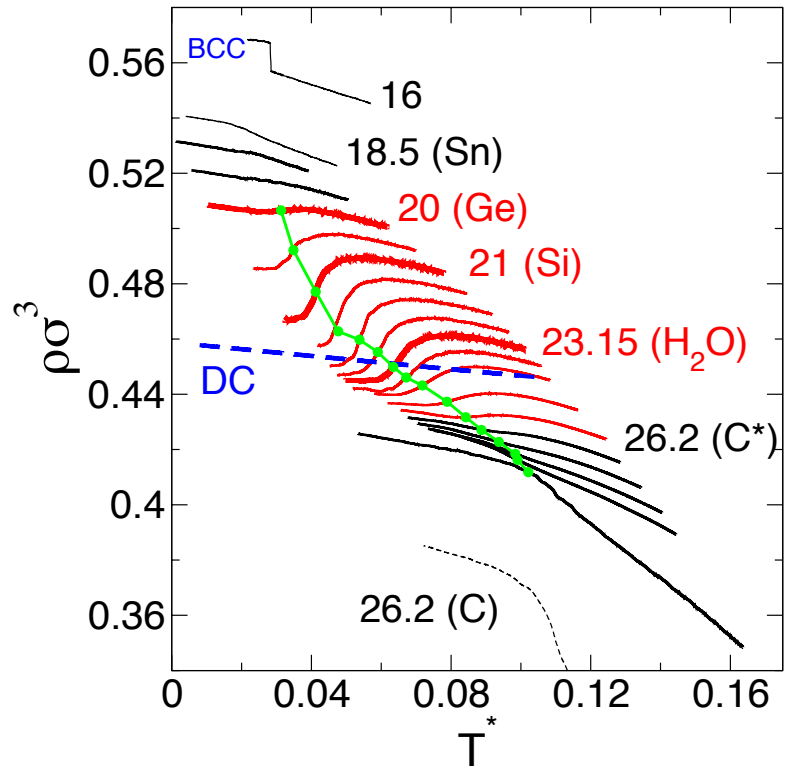

**Figure 2.** Phase diagram of the SW liquids in the $T$-$\lambda$ plane, for $p = 0$. The stable crystals have increasing coordination with decreasing tetrahedrality values: 4 for $\lambda > 18.75$ (cyan area), 6 for approximately $18.2 > \lambda > 18.75$ (green area) and 8 for $\lambda < 18.2$ (gray area). The white-filled diamond indicates the melting point of the liquid with $\lambda = 24.4$, for which the volume of melting is zero; more tetrahedral liquids are lighter than the tetrahedral crystal. The thick line in the $\lambda$ axis show the region where these liquids cannot be crystallized in the time scales of the simulations.[59] The loci of heat capacity maximum ($C_p^{max}$, shown as maroon line with white-filled squares) and maximum change in the structure of the liquid (maximum $df_4/dT$, indigo line with white-filled circles) are essentially identical. The liquids cannot be properly equilibrated below the line of $C_p^{max}$, thus values below that temperature may be affected by the cooling rate. The temperatures of maximum density (TMD, shown as red line with triangles up) and minimum density (TmD, shown as blue line with triangles down) meet at $\lambda = 26.4$. They would also meet at low $\lambda$ around 19, but the vitrification of the liquid prevents a precise determination of the low-$\lambda$ closure of the density anomaly loop. The temperature of maximum density (TMD) and minimum density (TmD) occur both in the supercooled region of the phase diagram. That is not the case for real water, for which the TMD is 4°C above the melting point

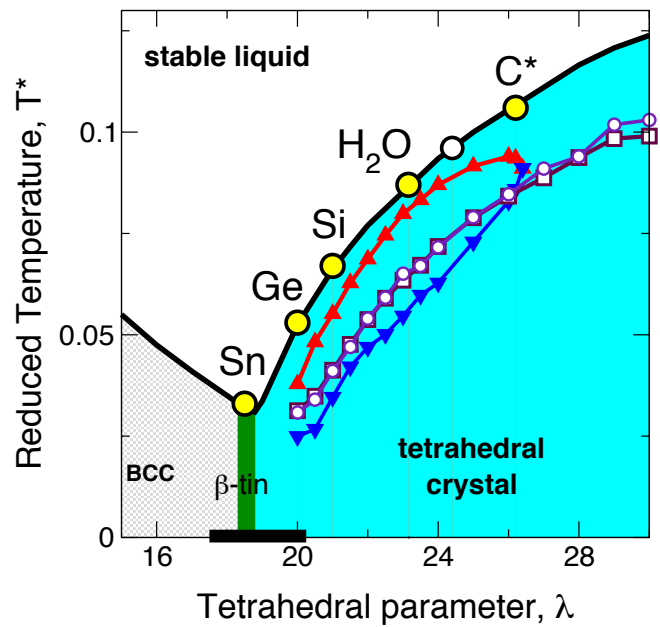

**Figure 3.** Variation of the pair correlation entropy $S_2$ as a function of reduced number density $\rho^*$ for different values of the tetrahedral parameter $\lambda$ (indicated in the graph). The vertical arrows represent the onset temperature for the pair entropy anomaly and correspond to 0.0379 ($\lambda$=20), 0.054 ($\lambda$=21) and 0.080 ($\lambda$=23.15). The horizontal arrows represent the maximum in heat capacity, which occur at 0.0326 ($\lambda$=20), 0.0422 ($\lambda$=21), 0.0651 ($\lambda$=23.15) and 0.0854 ($\lambda$=26.2). The curves are constructed from 512 particle simulations and are identical for 1K/ns and 5K/ns cooling protocols. Equilibration of the liquid at temperatures below the heat capacity maximum may be difficult due to high viscosity (e.g. Ge) or crystallization (e.g. C*).

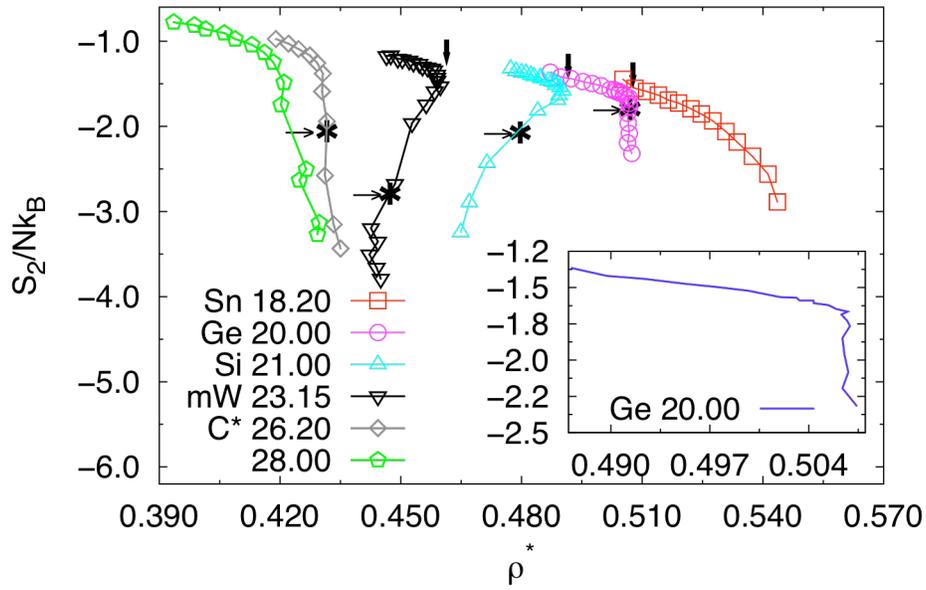

**Figure 4.** Variation of tetrahedral order $q_{tet}$ as a function of reduced temperature $T^* = Tk_B/\varepsilon$. The horizontal arrows represent the maximum in heat capacity, which occur at 0.0326 ($\lambda$=20), 0.0422 ($\lambda$=21), 0.0651 ($\lambda$=23.15) and 0.0854 ($\lambda$=26.2). The curves are constructed from 512 particle simulations and are identical for 1K/ns and 5K/ns cooling protocols. The glass transition temperature ($T_{glass}$) was defined to be the temperature for which no diffusive regime could be detected in the 1 ns period. The $T_{onset}$ temperature was defined as the temperature at which the onset of cooperative or caging effects in the dynamics are signaled by an inflexion point or a plateau region in the log-log plot of the MSD versus time.

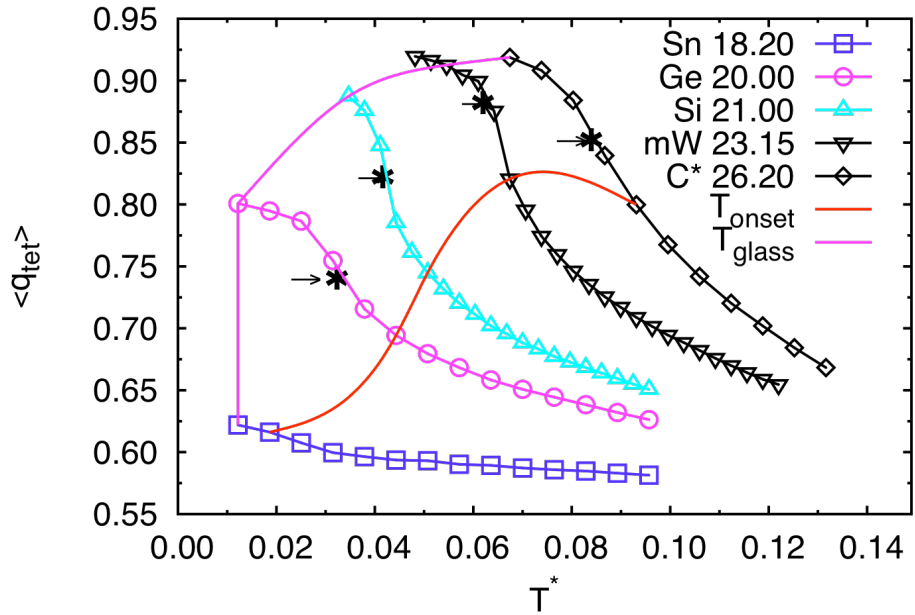

**Figure 5.** Tetrahedral order parameter distribution, $P(q_{tet})$, at different reduced temperatures $T^*$ (indicated in the figure) along $p=0$: for (a) $\lambda = 23.15$ (water) and (b) $\lambda = 26.2$ (carbon*). Arrows indicate the changes in the peak and shoulder features with increasing temperature.

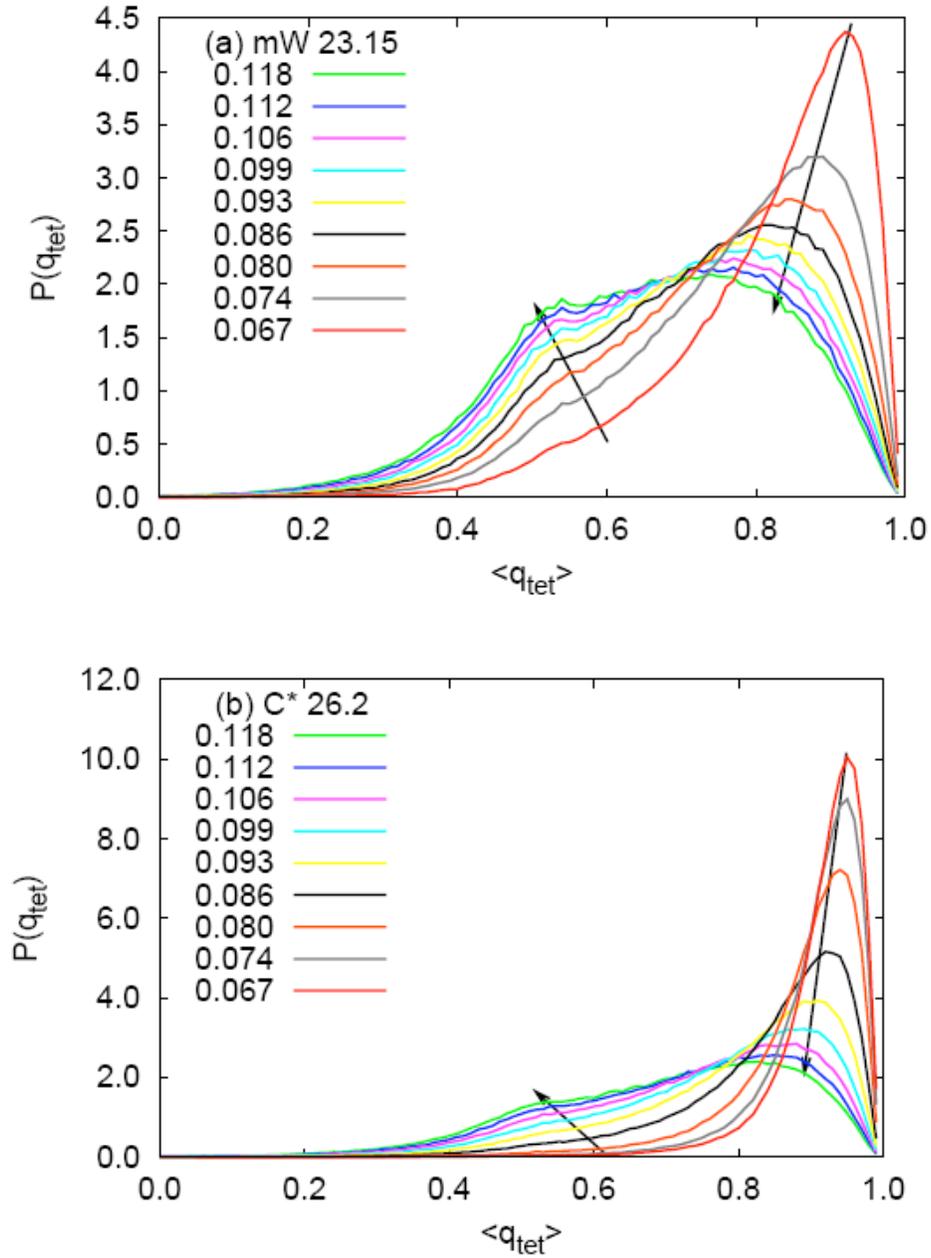

**Figure 6.** Reduced heat capacity $C_p/R$ as a function of reduced temperature $T^*$ for the family of SW liquids. $C_p$ was computed from quenching simulations of systems with 4,096 particles at a rate of 10 K/ns. Data is presented for liquids with tetrahedral parameters $\lambda$ = 30, 29, 28, 27, 26, 25, 24, 23.5, 23, 22,5, 22, 21,5, 21, 20.5, 20, 19.5 and 19. Liquids with $\lambda$ larger than 20.5 show a large increase of $C_p$ on cooling. Maximum value of the peak occurs around $\lambda$ = 22, an element intermediate between water and silicon.

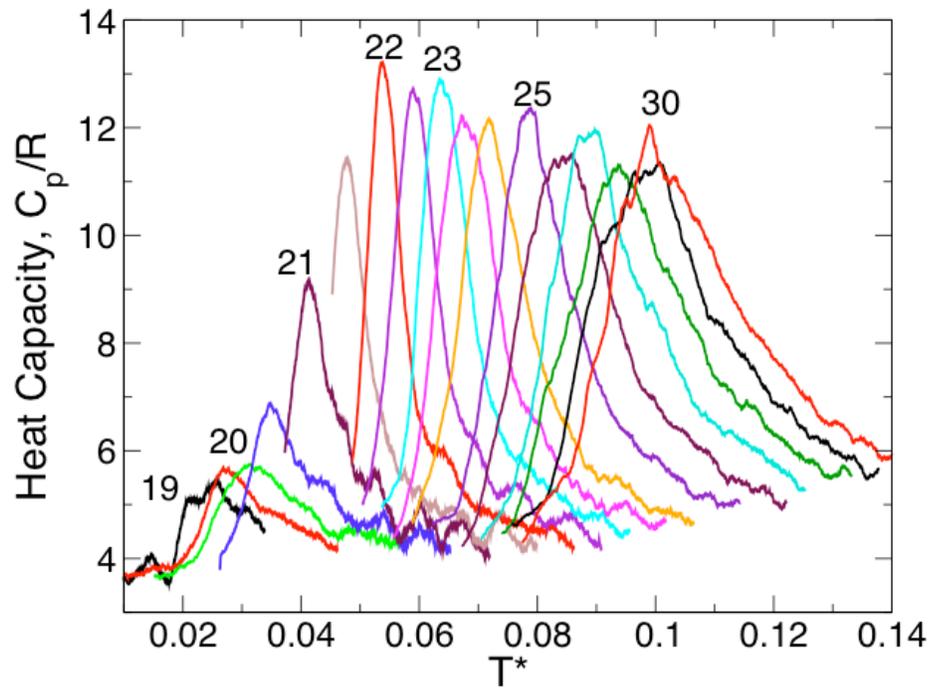